\documentclass[aps,twocolumn,superscriptaddress,10pt]{revtex4-2}

\usepackage{amsmath}    
\usepackage{graphicx}   
\usepackage{verbatim}   
\usepackage{color}      
\usepackage{hyperref}   
\begin{document}

\title{Analysis of electronic interactions in band- and Mott-insulating 1$T$-TaS$_2$ through variation of the electron count}
\title{Ultrafast Electronic Structure Engineering in  1$T$-TaS$_2$: \\Role of Doping and Amplitude Mode Dynamics}

\author{J. Jayabalan}
\affiliation{Faculty of Physics and CENIDE, University of Duisburg-Essen, 47048 Duisburg, Germany}
\author{Jiyu Chen}
\affiliation{Department of Physics, University of Fribourg, 1700 Fribourg, Switzerland}
\author{Laura Pätzold}
\affiliation{Institute of Theoretical Physics, University of Hamburg, 20355 Hamburg, Germany}
\author{Francesco Petocchi}
\affiliation{Department of Quantum Matter Physics, University of Geneva, 1211 Geneva, Switzerland}
\author{Florian K. Diekmann}
\affiliation{Institute of Experimental and Applied Physics, Kiel University, Olshausenstr. 40, 24098 Kiel, Germany}
\author{Negar Najafianpour}
\affiliation{Faculty of Physics and CENIDE, University of Duisburg-Essen, 47048 Duisburg, Germany}
\author{Ping Zhou}
\affiliation{Faculty of Physics and CENIDE, University of Duisburg-Essen, 47048 Duisburg, Germany}	
\author{Walter Schnelle}
\affiliation{Max Planck Institute for Chemical Physics of Solids, 01187 Dresden, Germany}
\author{Gesa-R. Siemann}
\affiliation{Department of Physics and Astronomy, Aarhus University, 8000 Aarhus C, Denmark}
\author{Philip Hofmann}
\affiliation{Department of Physics and Astronomy, Aarhus University, 8000 Aarhus C, Denmark}
\author{Kai Roßnagel}
\affiliation{Institute of Experimental and Applied Physics, Kiel University, Olshausenstr. 40, 24098 Kiel, Germany}
\affiliation{Ruprecht-Haensel-Laboratory, Deutsches Elektronen-Synchrotron DESY, 22607 Hamburg, Germany}
\author{Tim Wehling}
\affiliation{Institute of Theoretical Physics, University of Hamburg, 20355 Hamburg, Germany}
\author{Martin Eckstein}
\affiliation{Institute of Theoretical Physics, University of Hamburg, 20355 Hamburg, Germany}
\author{Philipp Werner}
\affiliation{Department of Physics, University of Fribourg, 1700 Fribourg, Switzerland}
\author{Uwe Bovensiepen}
\affiliation{Faculty of Physics and CENIDE, University of Duisburg-Essen, 47048 Duisburg, Germany}

\date{\today}

\begin{abstract}

In strongly correlated transition metal dichalcogenides, an intricate interplay of polaronic distortions, stacking arrangement, and electronic correlations determines the nature of the insulating state. Here, we study the response of the electronic structure to optical excitations to reveal the effect of chemical electron doping on this complex interplay. Transient changes in pristine and electron-doped 1$T$-TaS$_2$ are measured by femtosecond time-resolved photoelectron spectroscopy and compared to theoretical modeling based on non-equilibrium dynamical mean-field theory and density functional theory. The fine changes in the oscillatory signal of the charge density wave amplitude mode indicate phase-dependent modifications in the Coulomb interaction and the hopping. Furthermore, we find indications for an enhanced fraction of monolayers in the doped system. Our work demonstrates how the combination of time-resolved spectroscopy and advanced theoretical modeling provides insights into the physics of correlated transition metal dichalcogenides.

\end{abstract}

\maketitle

The properties of correlated materials are governed by many-body effects, which give rise to density waves, superconductivity, Mott-insulating ground states, and interesting nonequilibrium phenomena \cite{yang_review_2017,sipos2008mott,Giannetti2016,Sentef2021,Murakami_2023}. Among those compounds, the group 5 transition metal dichalcogenides (TMDC) are of particular interest due to their structural instabilities, charge density wave (CDW) formation, and Mott insulating character \cite{fazekas1979, perfetti2006time, sipos2008mott}.
A much-studied member of this family is 1$T$-TaS$_2$, which is metallic at temperatures $T$ above 540~K. As $T$ is lowered, a CDW is formed with increasing long range order. The CDW consists of individual Star-of-David shaped entities formed out of 13 Ta atoms each \cite{sipos2008mott}. At $T<350$~K these stars form clusters and a nearly commensurate CDW (NCCDW). Below 200~K a long range ordered, commensurate CDW develops into a ($\sqrt{13} \times \sqrt{13})R13.9^\circ$ superstructure of the Ta atoms.
In recent years, the early proposal of a Mott insulating ground state for 1$T$-TaS$_2$ by Fazekas and Tosatti \cite{fazekas1979} has been challenged. The stacking of individual layers was shown to be essential to understand the insulating character \cite{ritschel2015, ritschel2018, lee2019}. Surface and bulk contributions were considered \cite{Wang_2020, Dong_2023}, bilayer formation and Mott localization were disentangled, and it was shown that Mott insulating monolayers make up only about 20\% of the bulk \cite{butler2020, Hua_2025}.
Substitutional \cite{liu_2013,qiao2017} and hole doping \cite{zhang2023} have been investigated and coexisting metallic and insulating domains were reported.
On this basis, $1T$-TaS$_2$ has become an interesting material platform to study intertwined electronic and vibrational excitations  \cite{demsar2002, perfetti2006time, eichberger2010, dean2011, petersen2011, hellmann2012, butler2021, maklar2023}. Remarkably, the electronic structure changes associated with the CDW amplitude mode \cite{perfetti2006time, hellmann2012, avigo2018excitation, Dong_2023} are, so far, not understood microscopically.

In this work, we model the transient electronic structure change of the amplitude mode in $1T$-TaS$_2$ in real time and distinguish the response of mono- and bilayers in the parent and the electron-doped compound. Time-resolved photoemission spectra compare well with the developed theory. We establish, thereby, a phase-sensitive real-time observable of the dynamic electronic structure in strongly correlated 2D materials that provides additional insights into the stacking of mono- and bilayers.

\begin{figure}[t]
    \centering
    \includegraphics[width=0.99\columnwidth]{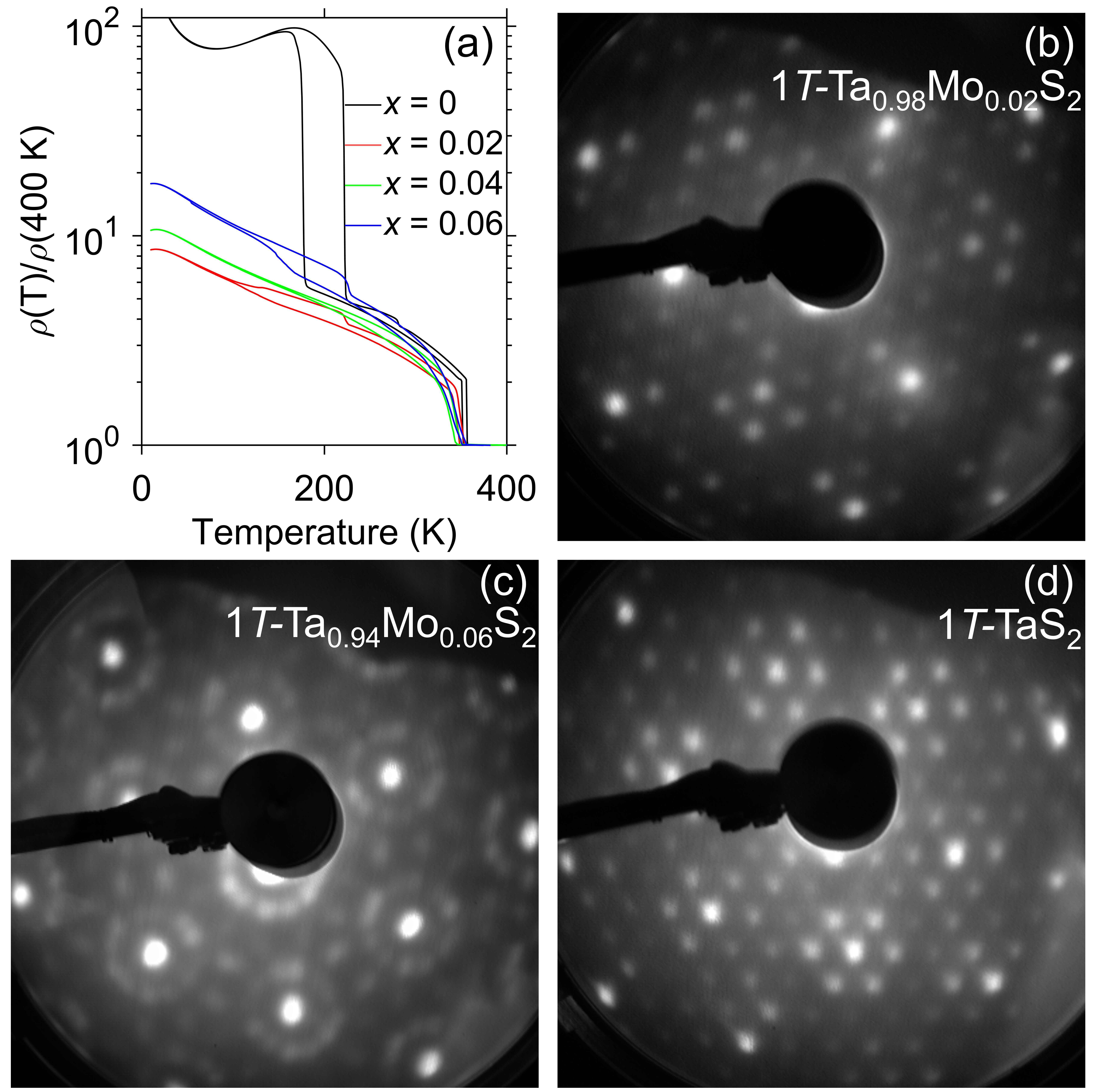}
    \caption{Panel (a) shows the temperature dependent electrical resistivity $\rho$ of 1$T$-Ta$_{(1-x)}$Mo$_x$S$_2$ at the indicated doping level $x$.  Panels (b-d) present LEED images
   for $x=0.02$ (b), 0.06 (c), and the parent compound $x=0$ (d), recorded at $T=30$~K with an electron kinetic energy of 123~eV.
    }
    \label{Fig_1}
\end{figure}

The temperature-dependent electric resistivity was measured for single crystals of 1$T$-Ta$_{(1-x)}$Mo$_x$S$_2$. The results are shown in Fig.~\ref{Fig_1}(a). The transition to the NCCDW phase prevails for all $x$ while the long-range order below $T=200$~K is strongly suppressed upon doping. The satellite features observed in LEED, see Fig.~\ref{Fig_1}(b-d) indicate that locally the Star-of-David (SOD) superstructure prevails, although at reduced intensity. Thus, Mo doping reduces the fraction containing the superstructure. For $x=0.06$ we identify more complex satellite features indicating two domains rotated with respect to each other by $30^ \circ \pm 4^\circ$. From these findings we conclude that electric conductivity channels persist in the presence of Mo doping at all temperatures. For further characterization, µ-ARPES studies were conducted. We found homogeneous crystal surfaces and no indication of dopant agglomeration, see the appendix~\ref{sec:µARPES}.

\begin{figure*}
    \centering
    \includegraphics[width=0.99\textwidth]{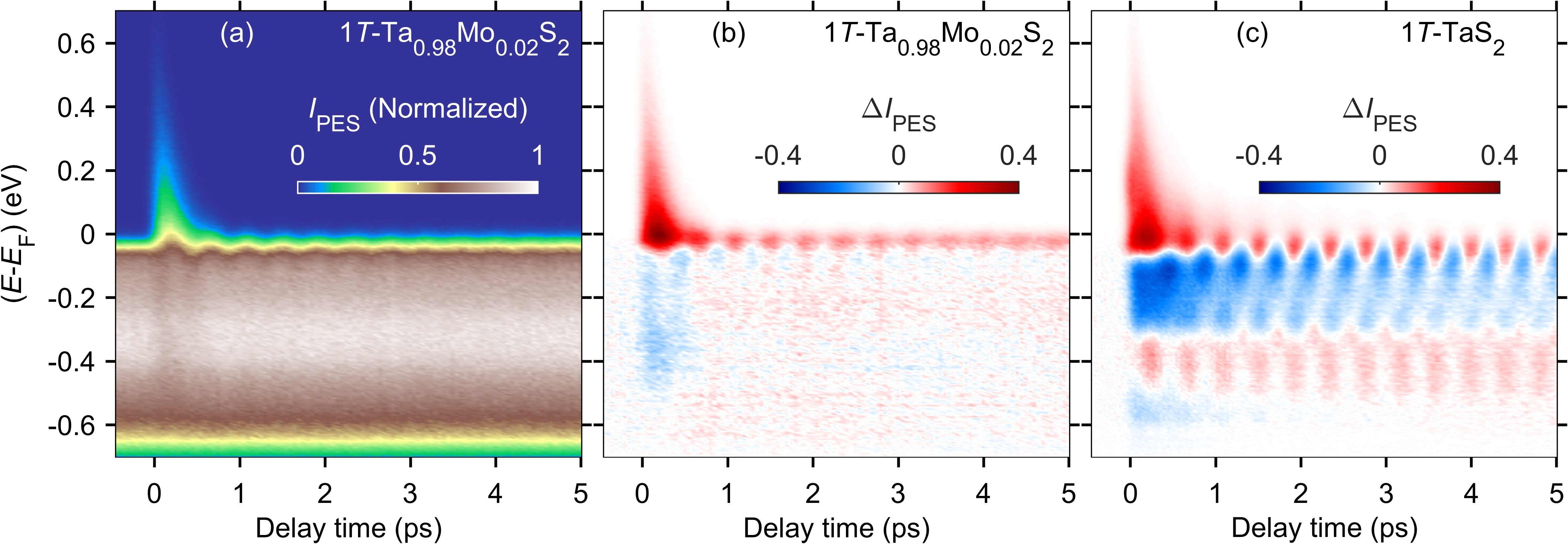}
    \caption{The time-dependent photoelectron intensity $I_{\mathrm{PES}}$ is depicted as a function of energy $E-E_{\mathrm{F}}$ and delay time $t$ in panel (a) for $T=30$~K and $x=0.02$ after normalization to the intensity maximum using a false color map. Panels (b,c) present the pump-induced change for $x=0.02$; $0$, respectively. Panels (a,b) show identical data; in (b) the static spectrum at $t<0$ was subtracted. Positive and negative changes are indicated in red and blue color, respectively. For $x=0$, full time-resolved spectra were previously published in Refs.  \cite{perfetti2006time, avigo2018excitation, Dong_2023}.}
    \label{Fig_2}
\end{figure*}

Femtosecond laser pulses with a fundamental photon energy of $h\nu=$1.51~eV and 50~fs pulse duration were used for pumping,  combined with probing by photoemission spectroscopy using the 4th harmonic at 6.04~eV and delay time $t$. The incident pump fluence was 0.2~mJcm$^{-2}$. The crystals were cleaved at 300~K. All measurements were performed at 30~K using $p$-polarized pulses. The kinetic energy of the photoelectrons emitted within an angle of $ \pm3^\circ$ to the normal surface was measured by electron time-of-flight spectroscopy \cite{sandhofer2014}. The Fermi energy $E_{\mathrm{F}}$ is determined by the high energy spectral cutoff without pumping at 300~K \cite{perfetti2006time}.

Time-dependent photoelectron emission spectra were taken for $1T$-Ta$_{(1-x)}$Mo$_{x}$S$_2$  with $x=0; 0.02; 0.04; 0.06$. Fig.~\ref{Fig_2}(a) shows the intensity for $x=0.02$ as a function of energy $E-E_{\mathrm{F}}$ and $t$. The spectrum at $t<0$ before excitation is 0.6~eV wide. At $t=0$ a slight intensity reduction below $E_{\mathrm{F}}$ and a clear increase above $E_{\mathrm{F}}$ represents the electronic response of the optical excitation. At delays $t<5$~ps, oscillations in the vicinity of $E_{\mathrm{F}}$ are observed. Their frequency is $\Omega=2.4$~THz and the oscillations dephase during 5~ps. This frequency represents the breathing of the Star-of-David and the CDW amplitude mode \cite{demsar2002, perfetti2006time, zong2018}. Subtraction of the spectrum at $t<0$ provides the dynamic spectral changes plotted for $x=0.02; 0$ in Fig.~\ref{Fig_2}(b,c). Very similar results were obtained for $x=0.06$.

We find pronounced transient spectral changes with characteristic differences for the parent and the electron doped compound. For $x=0$ they cover a wide range of 0.5~eV and exhibit a rich fine structure. For $x=0.02$ the transient changes are confined to 100~meV around $E_{\mathrm{F}}$ for $t>0.6$~ps. At earlier $t$, during the presence of hot holes, they reach to $E-E_{\mathrm{F}}=-0.5$~eV. Comparing $x=0$ and $x=0.02$ establishes that chemical doping by Mo confines the transient spectral changes, upon relaxation of hot charge carriers, to a narrow range around  $E_{\mathrm{F}}$. We attribute this behavior to a larger scattering phase space and enhanced screening induced by the increased electron density that originates from the additional valence electron in Mo compared to Ta.


\begin{figure}
    \centering
    \includegraphics[width=0.49\textwidth]{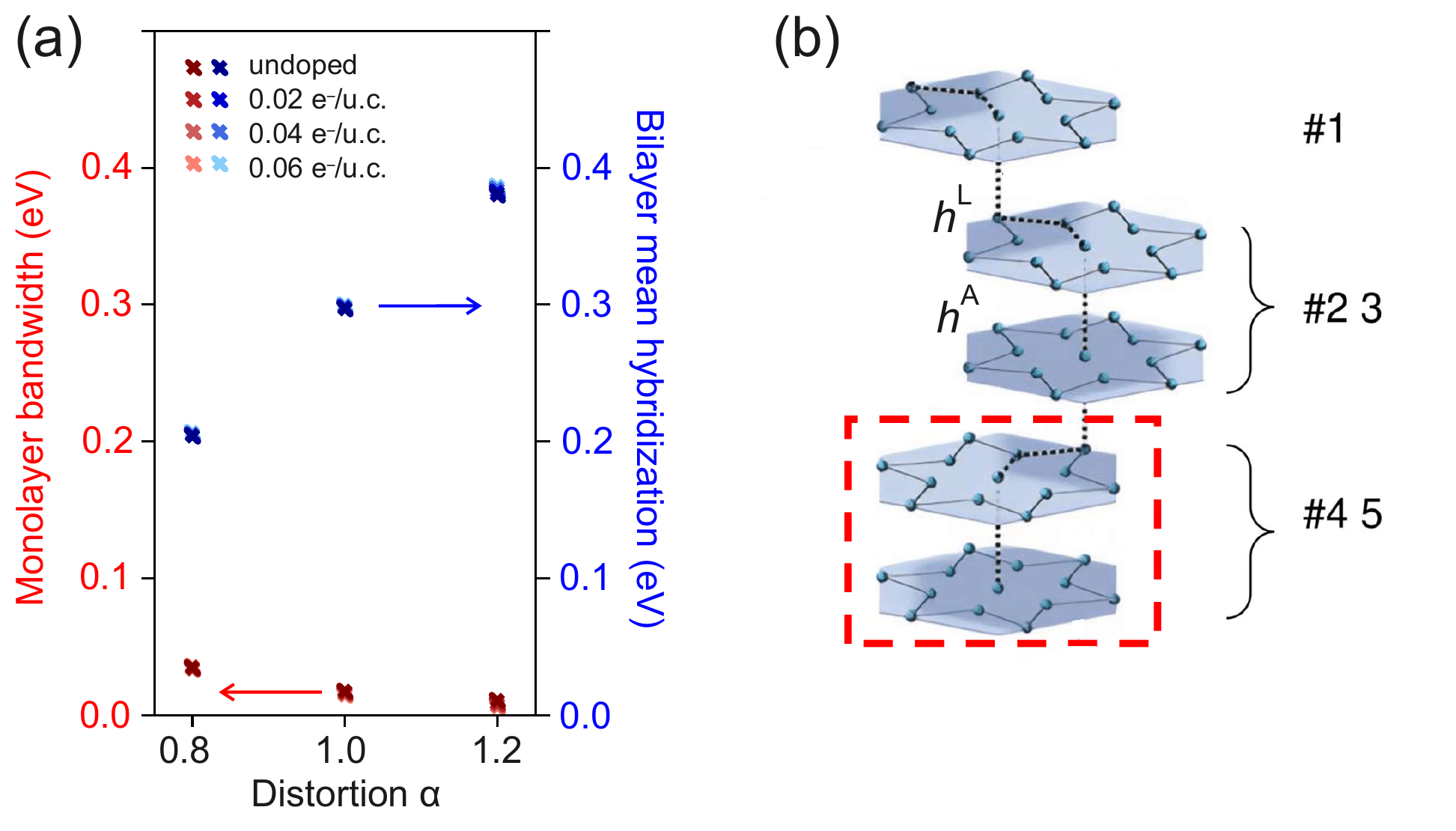}
    \caption{(a) Representative energy scales for mono- and bilayers of $1T$-TaS$_2$ calculated by DFT for the indicated doping. Left axis: Bandwidth for the monolayer single band as a function of the distortion $\alpha$ to simulate the influence of the breathing mode on the intralayer physics. Right axis: Calculated mean hybridization gap of the bilayer band structure for identical $\alpha$ representing the effect of the breathing mode on the interlayer physics. The interlayer energy scale is one order of magnitude larger than the intralayer one. (b) Stacking arrangements of the CDW ordered clusters located on different planes. The two-site DMFT cluster associated with a bilayer is indicated by the red box. Two hopping matrix elements $h^A$ and $h^L$ are considered, see text.
    }
    \label{Fig_3}
\end{figure}

To describe the changes in the electronic structure induced by the photoexcitation we combine density functional theory (DFT) calculations with nonequilibrium dynamical meanfield theory (DMFT) simulations. The results of the DFT calculations, which are summarized in the appendix~\ref{sec:calculation_details}, illustrate how the breathing mode of the SOD cluster changes the electronic structure at the well-known amplitude mode frequency of 2.4~THz \cite{demsar2002, perfetti2006time, stojchevska2014ultrafast}. We compare modulations of the electronic structure for a monolayer and a bilayer of $1T$-TaS$_2$, see Fig.~\ref{Fig:DFT_bands} (d-f) and (g-i) in the appendix. For monolayers (bilayers), the emergence of one (two) flat bands near $E_{\mathrm{F}}$ is found upon approaching the relaxed CDW structure characterized by a  distortion $\alpha=1$. $\alpha$ represents the scaled CDW distortion with the relaxed CDW structure corresponding to $\alpha=1$
and the high-temperature structure without CDW to $\alpha=0$. In the case of the bilayer, the two flat bands near $E_{\mathrm{F}}$ are split by a bonding-antibonding splitting due to interlayer hopping, which dominates over the in-plane kinetic energy contributions of the flat band. Upon reducing the CDW displacement amplitude by $40\%$, corresponding to $\alpha=0.6$, the flat bands become more dispersive in both cases and most strikingly, the bonding-antibonding splitting in the bilayer case is reduced. As visualized in Fig.~\ref{Fig_3}(a), the change in bonding-antibonding splitting is one order of magnitude larger than the effects on the in-plane kinetic energy contributions. This motivates the modulation of the interlayer hopping terms in the subsequent many-body treatment to mimic the dynamic effect of the Ta ion core vibration on the electronic structure.
At the same time, the oscillation of the cluster orbital associated with the flat band should lead to a modulation in the effective Coulomb repulsion.

To model the transient electronic structure we use a previously developed single-orbital multi-layer description of $1T$-TaS$_2$ in the CDW phase \cite{petocchi2022, petocchi2023}. The tight-binding Hamiltonian $\mathcal{H}_{\mathrm{SOD}}$ describing a single layer was obtained with DFT and maximally localized Wannier orbitals \cite{pasquier2022}. The resulting single-orbital model is supplemented with a local Hubbard repulsion to incorporate electronic correlations. To describe the stacking order, the neighboring layers are connected via two hopping amplitudes: $h^\text{A}=0.2$ eV for the intra-bilayer hybridization and $h^\text{L}=0.045$ eV for the hopping between two van-der-Waals bonded layers  \cite{petocchi2022, petocchi2023}, see Fig.~\ref{Fig_3}(b). The kinetic part of the Hamiltonian is
\begin{equation}
    \mathcal{H}_{ab}\left(\mathbf{R}\right)=\mathcal{H}_\text{SOD}
    \left(\mathbf{R}\right)\delta_{ab}-h_{ab}^{A}
\delta_{\mathbf{R},\mathbf{0}}
-h_{ab}^{L}
    \left(\mathbf{R}\right),
    \label{hamilt}
\end{equation}
where $\left\{a,b\right\}$ is a layer index and  $\mathbf{R}$ is the unit cell vector.
To study the time evolution after a photo-excitation pulse, we use the nonequilibrium generalization of DMFT~\cite{aoki2014} in combination with a non-crossing approximation (NCA)~\cite{keiter1971,eckstein2010} impurity solver. A simplified Bethe-lattice type self-consistency allows us to capture the essential correlation and hybridization effects by constructing an approximate real-space hybridization function~\cite{georges1996,werner2017,chen2024,chen2024a}, at the cost of losing the information on the dispersion relation. This approximation significantly reduces the computational effort and memory requirement, thus enabling multi-layer real-time simulations.

To fully capture the strong bilayer hybridization resulting from the intra-bilayer hopping $h^\text{A}$, we employ a two-site cluster DMFT \cite{lichtenstein2000}, where the two sites of a cluster represent the two planes of a bilayer (red box in Fig.~\ref{Fig_3}(b)). The local Hamiltonian for this cluster DMFT description of the bilayer has the form
\begin{align}
    \mathcal{H}_{i}\left(t\right)=&-\mu\left({n}_{i1}+{n}_{i2}\right) +U\left({n}_{i1\uparrow}{n}_{i1\downarrow}+{n}_{i2\uparrow} {n}_{i2\downarrow}\right) \\ \nonumber
    &-\sum_{\sigma}\left[h^{i}_{12}\left(t\right){c}_{i1\sigma}^{\dagger}{c}_{i2\sigma}+h.c.\right],
\end{align}
where $U$ is the local Hubbard interaction and $\mu$ is the chemical potential, which can be adjusted to realize the desired doping.

In the presence of the laser pulse, the hopping matrices become complex,
$h_{mn}(t) = h_{mn}(0)e^{i\phi_{mn}(t)}$,
where the Peierls phase $\phi_{mn}(t) =-\int_0^t d t^\prime\boldsymbol{E}(t^\prime)\cdot \boldsymbol{r}_{mn}$ is determined by the time integral of the vector product between the electric field of the laser $\boldsymbol{E}(t)$ and the real space displacement  $\boldsymbol{r}_{mn}$ between the centers of orbitals $m$ and $n$ \cite{aoki2014}. The DMFT loop starts from the atomic Green's functions $G_i$ of the different bilayers $i$ as initial guess. To construct the time-dependent hybridization function for bilayer $i$, we consider 18 DFT-derived  in-plane hoppings $\mathcal{H}_{i,\text{SOD}}(\mathbf{R})$ within bilayer $i$, and 3 weak inter-bilayer hoppings $h^\text{L}_{ij}(\mathbf{R}^\prime)$ connecting to polarons of bilayer $i$ with those in the adjacent bilayers $j$,
\begin{equation}
\begin{aligned}
    {\Delta}_{i}\left(t,t'\right)&=\sum_{\mathbf{R}\neq 0}\mathcal{H}_{\text{SOD},i}(\mathbf{R},t){G}_{i}(t,t')\mathcal{H}_{\text{SOD},i}(\mathbf{-R},t')\\
    & +\sum_{j=i\pm 1,\mathbf{R}^\prime} h^L_{ij}(\mathbf{R}^\prime,t){G}_{j}(t,t') {h}^L_{ji}(-\mathbf{R}^\prime,t'), \label{eq_delta}
\end{aligned}
\end{equation}
where we employed the Bethe-lattice-like connection to the cluster Green's functions for the bilayers $i$ and adjacent bilayers $j=i\pm 1$.
(The monolayer \#1 is treated as a bilayer with $h^A=0$.)
Using the hybridization function defined in Eq.~\eqref{eq_delta}  and the NCA solution for the cluster impurity problem, the bilayer Green's functions ${G}_{j}$ are self-consistently computed. 
In order to mimic the cooling effect from the coupling to degrees of freedom not included in the model, we locally add a boson bath in the form of a self-energy 
$(\Sigma_j)_{ab}^\text{bath}(t,t')=g^2\delta_{ab} D(t,t';\omega_0, T)G_{j,aa}(t,t')$,
which is diagonal in the layer index.
Here, $D(t,t';\omega_0,T)$
is
the bosonic Green's function \cite{aoki2014} for boson frequency $\omega_0=0.05$~eV and temperature $T=0.02$~eV/$k_{\mathrm{B}}$. We choose the coupling strength $g=0.08$~eV. To describe the effect of the breathing mode, we modulate $U$, $h^A$ and $h^L$ by 5\%, with the frequency $\Omega=2.4$~THz.


\begin{figure*}
    \centering
    \includegraphics[width=0.99\textwidth]{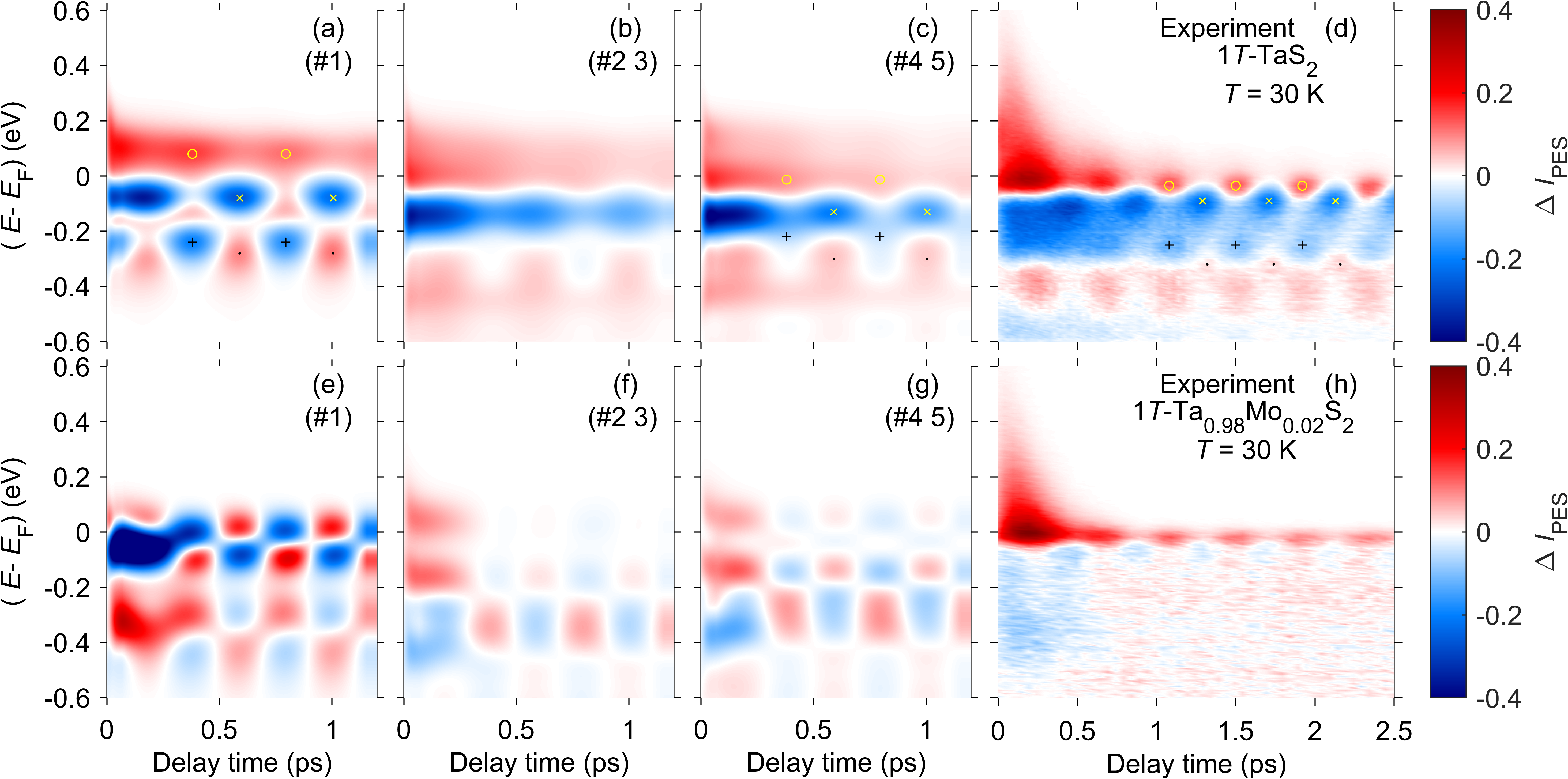}
    \caption{False color representation of the calculated (a-c, e-g) and measured (d,h) transient spectral changes for the monolayer at the surface \#1, the 1\textsuperscript{st} bilayer \#2 3, and the 2\textsuperscript{nd} bilayer \#4 5. The changes in the calculated spectral functions reflect the relative changes between the mono- and bilayers. Experimental results are replotted from Fig.~\ref{Fig_2}. The top panels represent $x=0$, the bottom ones $x=0.02$.}
	\label{Fig_4}
\end{figure*}

Figure~\ref{Fig_4} presents the calculated time-dependent changes in the spectral function for $x=0$ in panels (a-c) and for $x=0.02$ in (e-g) for the monolayer \#1 at the surface (a,e), for the first bilayer \#23 (b,f), and the second bilayer \#45 (c,g). These results compare reasonably well with the experimental results that are replotted in panels (d,h). Comparing experimental and theoretical results allows a discussion of the bi- and monolayer contributions in the cleaved crystal that are probed in the experiment. In our modeling we find for $x=0$ above $E_{\mathrm{F}}$ decaying hot electron contributions that are strong and contain a weakly oscillating contribution. For $x=0.02$ all calculated hot electron contributions are weaker than for $x=0$. At energies  $E-E_{\mathrm{F}}<0.1$~eV oscillatory signatures of the amplitude mode are present in all cases, although their strength varies.

In the following we concentrate on the monolayer and second bilayer since they exhibit clear phase and energy dependent changes. In the experiment, the negative changes for $x=0$ are very pronounced at $0<E-E_{\mathrm{F}}<-0.3$~eV and match well with the calculations in this energy interval for the monolayer. Calculations and experiment also match well in this energy range for positive changes $\Delta I_{\mathrm{PES}}$. At the phase characterized by the maximal $|\Delta I_{\mathrm{PES}}<0|$ at high energy, these features are indicated in panels (a,d) by a yellow ``$\times$" symbol, $\Delta I_{\mathrm{PES}}>0$ indicated by a black ``$\cdot$" appears at lower energy. At the opposite phase with maximal $|\Delta I_{\mathrm{PES}}<0|$ at the lowest energy, indicated by a black ``$+$" symbol, $\Delta I_{\mathrm{PES}}>0$, marked by a yellow ``$\circ$" symbol, appears at high energy. This oscillatory fine structure is found in experiment and theory for the monolayer,
although at short times (when doublon/holon populations exist in the Hubbard bands), the main intensity is above $E_\mathrm{F}$.
Considering that differential signals are discussed, 
these oscillations
around a central energy of $E-E_{\mathrm{F}}=-180$~meV 
indicate a modulation of
the Hubbard $U$.
The fact that positive and negative changes $\Delta I_{\mathrm{PES}}$ occur at opposite phases at different energy implies that the line profile and, thus, the hopping rate varies within the oscillation cycle. In the experiment the maxima and minima in $\Delta I_{\mathrm{PES}}$ occur at 30~meV energy difference on the higher and lower energy spectral sides. In the simulations, the results differ slightly for the higher and lower energy spectral sides, probably due to the large spectral weight obtained in the calculation above $E_{\mathrm{F}}$. However, as an order of magnitude estimation we conclude on substantial changes in the hopping rates of several 10~meV.

Considering that photoemission spectroscopy is a surface sensitive technique the question of a monolayer or bilayer surface termination is relevant. Comparing panels (c,d) in Fig.~\ref{Fig_4},  we see that also the bilayer contribution can qualitatively explain the experimental signal. In particular the time-dependent signature of the simulated bilayer behavior directly at $E_{\mathrm{F}}$ matches well with the experiment, since the photo-doping of bilayers generates in-gap states \cite{petocchi2023}. The bilayer signal at lower energy however appears to be weaker than in the experiment.


Turning to the results for the doped crystals, we find that a periodic intensity change is found at $E_{\mathrm{F}}$ for the bilayer
and for the monolayer simulation, as well as in the experiment. The changes for the bilayer are however much weaker than for the monolayer and do not vary in the sign of $\Delta I$ across $E_{\mathrm{F}}$,  which is clearly inconsistent with the experiment.
In the simulation, the strong modification of the spectral weight at $E_F$ reflects a periodic modulation of a narrow quasi-particle band.
We, therefore, conclude that for the doped case the monolayer contribution dominates.

A recent x-ray scattering study \cite{Hua_2025} reported a ratio of bilayers to monolayers in the bulk of pristine 1$T$-TaS$_2$ of 2:1, i.e. 20\% of the layers are monolayers. Our results for $x=0$ are consistent with a significant bilayer contribution, as discussed above.
For the doped compound, the calculated monolayer spectral function shows a much better agreement with the experimental result than the bilayer spectral function. This indicates that the contribution of monolayers in the doped system is higher than in the undoped compound, at least near the surface probed by ARPES. Apparently, the doping not only reduces the CDW fraction as concluded from LEED, see Fig.~\ref{Fig_1}, but also hinders the formation of bilayers. Since the stacking order influences the physics in profound ways, this finding is significant. It suggests that in addition to the appearance of metallic channels between polaronically distorted domains, chemical doping triggers a restructuring of the layered crystal and fundamentally changes the nature of electronic correlations.


Having drawn this conclusion we note that the match of the oscillatory spectral fine structure between experiment and theory is not perfect. The spectral changes calculated for the doped case reach considerably higher energies above $E_{\mathrm{F}}$ than the experimental spectra. We assign this difference to the construction of the model, which considers only the low-energy band(s).
In the real material, additional bands are present, and may play a role in the excitation and relaxation process.
Furthermore, in the experiment for $x=0$, we observe at $E-E_{\mathrm{F}}=-0.4$~eV an additional oscillatory signature with $\Delta I>0$. This energy agrees well with the CDW energy gap at the Brillouin zone boundary at the M-point detected in static angle-resolved PES (ARPES) \cite{petersen2011, hellmann2012}. In time-resolved ARPES the CDW gap oscillates at the amplitude mode frequency, in agreement with the $\Omega=2.4$~THz identified in the present work using $h\nu=6$~eV. At this low photon energy, ARPES does not reach the M-point. Moreover, we detect the photoelectron in normal emission at $\Gamma$. Following Petersen {\it et al.} \cite{petersen2011}, we assign the oscillation at $-0.4$~eV to the amplitudon whose excitation coherently modifies the scattering probability leading to  backfolding from M to $\Gamma$. Our modeling does not take the long-range ordered crystal lattice into account and lacks this backscattered feature.

We presented phase-resolved microscopic insights into dynamic electronic structure changes in response to the amplitude mode of the charge density wave in parent and Mo-doped $1T$-TaS$_2$. Comparing results of time-resolved photoelectron spectroscopy with real-time dynamical mean field simulations combined with parameter modulations inspired by density functional calculations, we resolved periodic variations in the Coulomb repulsion and the hopping rate. Furthermore, we were able to draw conclusions on the stacking in the material. While for the parent compound we favor a coexistence of mono- and bilayers near the surface, the Mo doping increases the monolayer fraction and, along with this restructuring, the effect of strong electronic correlations. Such an analysis of dynamic changes of the electronic structure combined with electronic structure engineering through accessible control parameters provides valuable insights into the fundamental electronic interactions governing complex materials with considerable potential for future analysis.


\begin{acknowledgments}
The authors acknowledge the funding by DFG through QUAST-FOR5249 and Project No. 278162697 - SFB 1242. The DMFT calculations were run on the beo05 cluster at the University of Fribourg. L.P. and T.W. gratefully acknowledge the computing time made available to them on the high-performance computer "Lise" at the NHR Center NHR@ZIB under the project hhp00063. This center is jointly supported by the Federal Ministry of Education and Research and the state governments participating in the NHR (\url{www.nhr-verein.de/unsere-partner}).

\end{acknowledgments}

\appendix

\section{Analysis of sample homogeneity}
\label{sec:µARPES}
\begin{figure*}[t]
    \centering
    \includegraphics[width=0.99\textwidth]{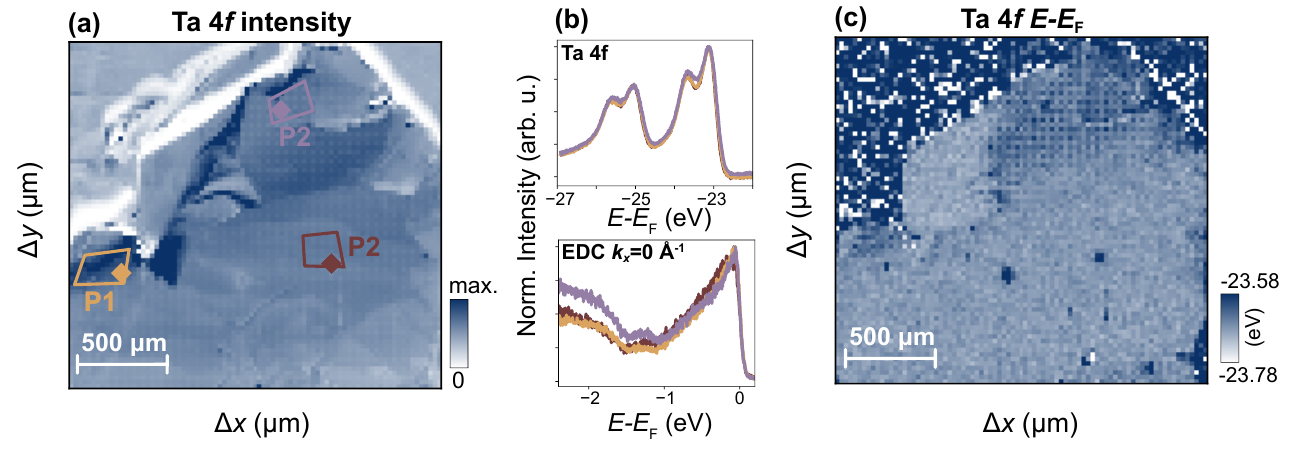}
    \caption{(a)  Spatially resolved ARPES measurements scanning over the Ta 4$f$ core levels using a photon energy of h$\nu = 80$~eV and LH-polarized light. (b) Top: Extracted core levels from the map shown in (a), integrated over small regions around three different probing spots. Bottom: Extracted EDCs at the Gamma point ($k_x$ = 0~\AA$^{-1}$) from fixed scans measured at the same three positions, indicated by the colored markers in (a) using a photon energy of h$\nu = 50$~eV. (c) Variations in the Ta 4$f$ binding energy of the third peak component indicated by the arrow in (b), showing only small variations in the binding energy within the topmost part of the sample.}
    \label{Fig:muARPES}
\end{figure*}

In Fig.~\ref{Fig:muARPES}, we present spatially resolved ARPES measurements over a large area of a 6\% Mo-doped TaS$_2$ sample using a step size of 25~$\mu$m. The raw variations in the Ta 4$f$ core-level intensity across the sample are shown in Fig.~\ref{Fig:muARPES}(a). Normalized core-level spectra extracted from three distinct regions indicated by colored areas in Fig.~\ref{Fig:muARPES}(a) are plotted in the top of Fig.~\ref{Fig:muARPES}(b), demonstrating no significant variations across the sample surface. To further confirm the uniformity of the sample, we investigated variations in the valence band electronic structure at different positions marked by colored symbols in Fig.~\ref{Fig:muARPES}(a), using a photon energy of 50~eV. High statistics measurements of the Valence band spectra at normal emission, represented by EDCs at the $\Gamma$ point using a 0.06~{\AA}$^{-1}$ integration range (bottom panel of Fig.~\ref{Fig:muARPES}(b)), indicate that observed spectral differences arise primarily due to slight angular offsets between measurement positions. Additionally, the homogeneity was evaluated by fitting the Ta 4$f$ core-level spectra at each spatial point of the map using four peaks convoluted with a Gaussian function and a linear background. The resulting spatial variation of the binding energy for the third peak component (indicated by the gray arrow in Fig.~\ref{Fig:muARPES}(b)) is plotted in Fig.~\ref{Fig:muARPES}(c), revealing only minor variations in the uppermost region of the sample.

\section{DFT band structure calculations}
\label{sec:calculation_details}

DFT calculations were performed with \textsc{Quantum ESPRESSO}~\cite{Giannozzi2009} using the PBE approximation for the exchange-correlation potential~\cite{Perdew1996} and norm-conserving pseudopotentials \cite{Hartwigsen1998}.
Plane waves until an energy cutoff of \(100\thinspace\text{Ry}\) were included and a Fermi-Dirac-type smearing of \(0.005\thinspace\text{Ry}\) was imposed.
We used a lattice constant of \(3.43\thinspace\)\AA{} and a $6\times 6\times 1$ $\mathbf k$-point grid for the $\sqrt{13} \times \sqrt{13}$ supercell. A vacuum of approximately \(12\thinspace\)\AA{}~was included above the monolayer/bilayer.
For calculations of the doped system, we considered a doping of \(0.02/0.04/0.06\thinspace\) e\(^{-}\) per TaS$_2$ unit cell, which corres\-ponds to \(0.26/0.52/0.78\thinspace\) e\(^{-}\) per $\sqrt{13}\times\sqrt{13}$ supercell.

\begin{figure*}[t]
    \centering
    \includegraphics[width=0.99\textwidth]{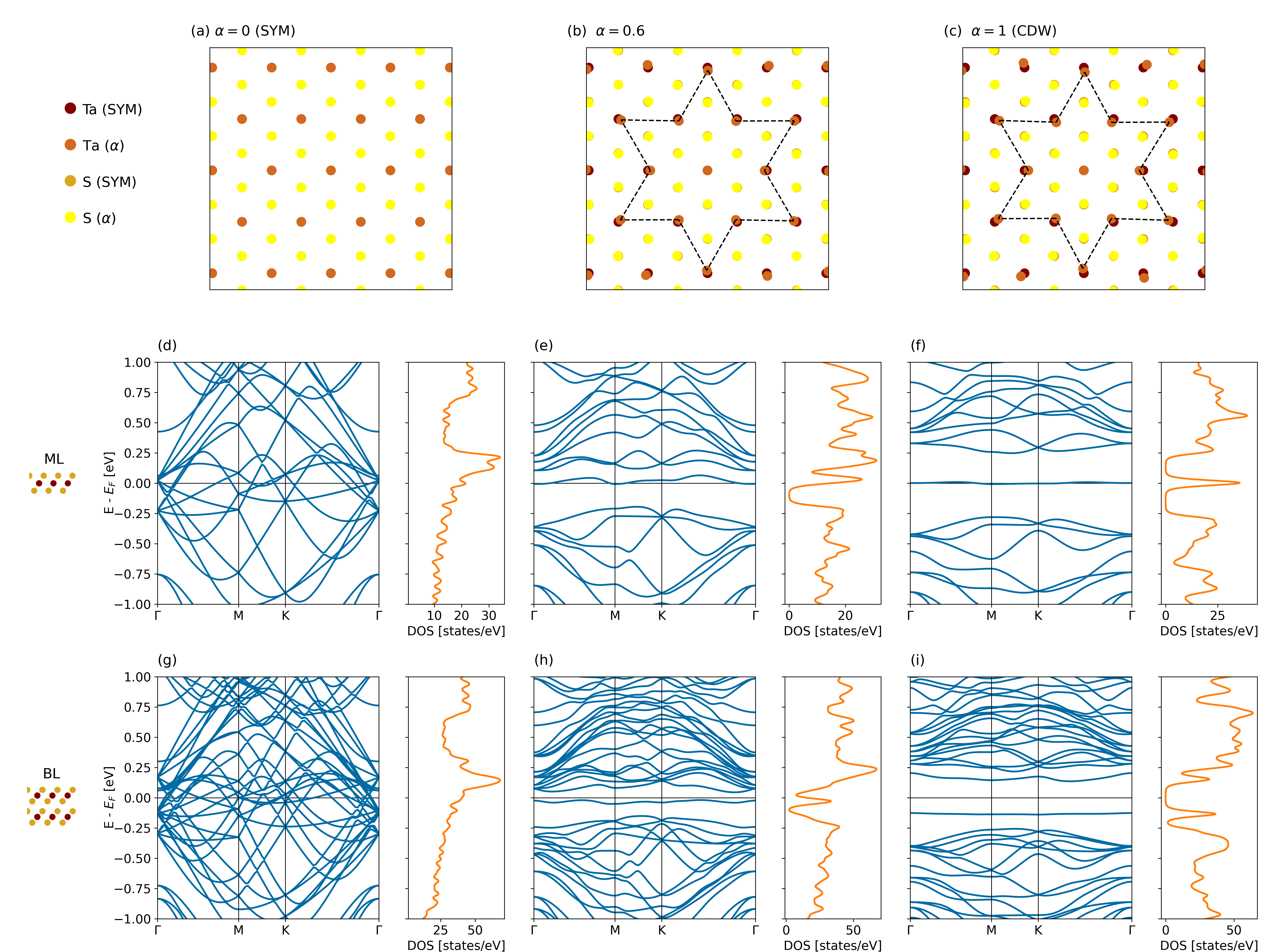}
    \caption{DFT calculations for monolayer and bilayer 1$T$-TaS$_2$. (a)-(c)~Zoom into the $\sqrt{13}\times\sqrt{13}$ supercell with a lattice constant of \(3.43\thinspace\)\AA{} for different percentages of displacement towards the SOD distortion, from (a)~the symmetric phase $\alpha=0$ over (b)~$\alpha=0.6$ to (c)~the SOD CDW $\alpha=1$. As a guide to the eye, the star shape of the CDW is marked via dashed lines. (d)-(f)~Non-interacting band structure (blue) and density of states (DOS, orange) of the 1$T$-TaS$_2$ monolayer obtained via DFT for the different percentages of distortion $\alpha$ from (a)-(c). The emergence of a single band around the Fermi level can be seen. (g)-(i)~Non-interacting band structure (blue) and density of states (DOS, orange) of the 1$T$-TaS$_2$ bilayer obtained via DFT for the different percentages of distortion $\alpha$ from (a)-(c).}
    \label{Fig:DFT_bands}
\end{figure*}

%

\end{document}